\documentclass[aps,prb,twocolumn,superscriptaddress]{revtex4-2}

\usepackage{float}
\usepackage{amsmath}
\usepackage{mathtools} 
\usepackage{amssymb}
\usepackage{amsthm}
\usepackage{bm} 
\usepackage{upgreek} 
\usepackage{xcolor}
\usepackage{graphicx}
\usepackage{subfigure}
\usepackage{epstopdf}
\usepackage{ulem}
\usepackage[colorlinks,linkcolor=blue,anchorcolor=blue,citecolor=blue,urlcolor=blue]{hyperref}

\newcommand{\nn}{{\nonumber}}

\newcommand{\av}[1]{\left\langle #1 \right\rangle}
\newcommand{\md}{\mathrm{d}}

\newcommand{\w}{\omega}

\begin{document}
\title{A functional renormalization group study of the two dimensional Su-Schrieffer-Heeger-Hubbard model}

\author{Qing-Geng Yang}
\affiliation{National Laboratory of Solid State Microstructures $\&$ School of Physics, Nanjing University, Nanjing 210093, China}

\author{Da Wang}\email{dawang@nju.edu.cn}
\affiliation{National Laboratory of Solid State Microstructures $\&$ School of Physics, Nanjing University, Nanjing 210093, China}
\affiliation{Collaborative Innovation Center of Advanced Microstructures, Nanjing University, Nanjing 210093, China}

\author{Qiang-Hua Wang} \email{qhwang@nju.edu.cn}
\affiliation{National Laboratory of Solid State Microstructures $\&$ School of Physics, Nanjing University, Nanjing 210093, China}
\affiliation{Collaborative Innovation Center of Advanced Microstructures, Nanjing University, Nanjing 210093, China}

\begin{abstract}
We study the Hubbard model on the square lattice coupled in addition to the optical Su-Schrieffer-Heeger (SSH) phonons, using the singular-mode functional renormalization group method.
At half-filling and in the absence of the Hubbard interaction $U$, we find the degenerate spin-density-wave (SDW)/charge-density-wave (CDW)/s-wave superconductivity (sSC) state at smaller electron-phonon coupling strength $\lambda$ and higher phonon frequency $\omega$, and the valence bond solid (VBS) state at larger $\lambda$ and lower $\omega$.
After switching on a positive $U$, the VBS state is suppressed, while the SDW state is enhanced.
At finite doping, the SSH phonon is found to favor sSC at $U=0$.
With increasing positive $U$, we find d-wave superconductivity (dSC) and incommensurate SDW states.
In a narrow window of moderate $U$ and $\lambda$, we also find the incommensurate VBS state. The sSC and dSC here can be naturally related to the CDW and SDW fluctuations, both of which can be triggered by the SSH phonons. In contrast, the repulsive interaction $U$ enhances SDW but suppresses CDW fluctuations. Our results at half filling are consistent with quantum Monte Carlo (QMC), and provide insights at finite doping where QMC may suffer from the minus sign problem.
\end{abstract}
\maketitle

\section{INTRODUCTION}
The electron-phonon interaction (EPI) is one of the fundamental interactions in condensed matter physics.
It plays important roles in many quantum phenomena, such as the Peierls transition  \cite{peierls1991more}, topological solitons \cite{su1979solitons}, and Bardeen-Cooper-Schrieffer (BCS) superconductivity \cite{bardeen1957theory}.
Over the past few decades, a growing number of experiments suggest that the EPI may be necessary in understanding various novel phenomena in strongly correlated materials \cite{lanzara2001evidence, cuk2005review, lu2012angle, chen2021anomalously, wang2012interface, zhong2016nodeless, lee2006interplay, gadermaier2010electron, song2019evidence, zhang2019enhanced, zhao2018direct, peng2020picoscale, pintschovius1997observation}, which has stimulated many theoretical studies in this direction \cite{mishchenko2004electron, wang2015phonon, liu2016giant, esterlis2018breakdown, costa2018phonon, li2019enhancement, zhang2019charge, chen2019charge, batrouni2019langevin, li2019electron, gao2020strong, cohen2020langevin, feng2020interplay, chen2020enhanced, wang2021phonon}.
Among different types of EPIs, the Su-Schrieffer-Heeger (SSH) model was first introduced to study the physics of {\it trans} polyacetylene \cite{su1979solitons,heeger1988solitons}.
Interestingly, the original acoustic SSH phonon model is found to give similar results to the optical SSH phonon model \cite{fradkin1983phase,weber2015excitation} since the dominant scattering process mainly involves the phonon modes near zone boundary rather than the zone center.
In the new century, the SSH model gets renewed attention because it can be taken as the simplest one-dimensional (1D) topological insulator \cite{qi2011topological}, can be mapped to a 1D topological superconductor \cite{kitaev2001unpaired}, and has been realized in optical lattices \cite{atala2013direct, meier2016observation, lohse2016thouless}.

Recently, the studies of the 1D SSH model have been extended to two dimensions (2D) \cite{feldbacher2003coexistence, xing2021quantum, cai2021antiferromagnetism, cai2021robustness, gotze, feng2022phase, yirga2022phonon}.
The optical SSH model on the square lattice without or with the Hubbard interaction $U$ has been carefully studied by the quantum Monte Carlo (QMC) method at half filling, where the negative sign problem is absent.
In the anti-adiabatic limit (with the phonon frequency $\omega\to
\infty$), the EPI induces effectively an instantaneous bond-wise interaction and has been studied in an early work \cite{feldbacher2003coexistence}.
In the absence of $U$, it is found that a degenerate spin-density-wave (SDW)/charge-density-wave (CDW)/s-wave superconductivity (sSC) state is present for any nonzero EPI strength. Both the CDW and SDW states have momentum $\0Q=(\pi,\pi)$, while the sSC is at momentum zero (or uniform).
The degeneracy is protected by the underlying spin-SU(2), charge-SU(2), and a Z$_2$ symmetry that treansforms between the two sectors \cite{yang1990so}. A spontaneous symmetry breaking in the internal space is needed for the system to settle down in a specific one of the degenerate states. After a repulsive $U$ is turned on, the Z$_2$ symmetry is broken and the SDW state is the only instability in the anti-adiabatic limit \cite{feldbacher2003coexistence}.

For finite phonon frequency and $U=0$, the degenerate SDW/CDW/sSC order still exist for weak EPI, but a strong enough EPI drives the system into a charge-bond-order state with ordering momentum $\0Q=(\pi,\pi)$, which is also referred to as valence-bond-solid (VBS) or bond-order-wave in some literature \cite{cai2021antiferromagnetism, xing2021quantum}. In fact, this can also be regarded as a generalized CDW state for charges defined on bonds rather than on sites. For comparison, in the 1D SSH model,  any nonzero EPI can induce the VBS, referred to as dimerization \cite{fradkin1983phase, sengupta2003peierls, bakrim2015nature}, as a result of the Peierls instability \cite{peierls1991more}.
The effect of a finite Hubbard $U$ has also been considered very recently \cite{cai2021robustness, gotze, feng2022phase, yirga2022phonon}.
Since $U$ breaks the Z$_2$ symmetry, a positive $U$ enhances the SDW in the spin channel while suppresses the CDW/sSC in the charge channel, providing a new avenue to explore the transition between SDW and VBS \cite{dqcp, gotze}.

These recent studies uncover rich physics of the 2D SSH-Hubbard model.
But most of these studies are restricted to half-filling due to the negative sign problem in QMC at finite doping (and for positive $U$).
In this work, we investigate the optical SSH-Hubbard model with the singular-mode functional renormalization group (SMFRG) method \cite{wang2012functional, xiang2012high, wang2013competing}, which treats all electronic orders on an equal footing for all cases of interactions and doping levels under concern.
At half-filling with $U=0$, we find the VBS state at larger $\lambda$ and lower phonon frequency $\omega$, and we obtain the degenerate SDW/CDW/sSC states at lower $\lambda$ and higher $\omega$. We also point out an interesting accidental degeneracy between the VBS and the Pomeranchuk state for $\omega=0$ and $\lambda\to 0$ (in the thermodynamic limit). After switching on the repulsive $U$, the SDW/CDW/sSC states are further split, leaving only the SDW as the dominant instability.
Next, we study the doped case (e.g., with $\av{n}=0.85$ electrons per site).
The SSH phonon always induces sSC in the absence of $U$, and this can be traced back to the CDW fluctuations the EPI triggers. With increasing positive $U$, the d-wave SC (dSC) and incommensurate SDW states appear successively. The dSC can be traced back to the antiferromagnetic SDW fluctuations. We also find incommensurate VBS state in a narrow window of moderate $U$ and $\lambda$.

The rest of this paper is organized as follows. In Sec.\ref{sec:model}, we describe the model mainly focusing on the symmetries and different ordering candidates, and briefly sketch the SMFRG method. Then we present and discuss the SMFRG results in
Sec.~\ref{sec:results}, followed by a summary of this work in Sec.~\ref{sec:summary}.

\section{MODEL AND METHOD} \label{sec:model}

We study the SSH-Hubbard model on the square lattice described by the Hamiltonian
\begin{align}
H=&  -t\sum_{\av{ij}\sigma}(c^\dagger_{i\sigma}c_{j\sigma}+{\rm H.c.}) - \mu \sum_{i}n_{i}
+ \frac{U}{2} \sum_{i}(n_i-1)^2\nonumber \\
& +\frac{g}{\sqrt{2M\omega}} \sum_{\av{ij}\sigma}(b_{ij}+b_{ij}^\dag)(c^\dagger_{i\sigma}c_{j\sigma}+{\rm H.c.})\nonumber\\
&
+ \w\sum_{\av{ij}}\left( b_{ij}^\dag b_{ij}+\frac12\right),
\end{align}
where $t$ is the hopping integral on nearest neighbor bond $\langle ij \rangle$, $c^\dagger_{i\sigma}$ creates an electron at site $i$ with spin $\sigma$,
$n_i=\sum_{\sigma} c_{i\sigma}^\dagger c_{i\sigma}$ is the local electron density, $\mu$ is the chemical potential,
and $U$ is the on-site Hubbard interaction.
We set $t=1$ as the unit of energy henceforth.
The optical SSH phonon is created by $b_{ij}^\dag$ on each bond $\langle ij \rangle$ with frequency $\w$.
These bond phonons couple to the electrons in the bond charge channel through the $g$-term, where $M$ denotes the effective mass for the optical vibration.

The model enjoys the natural global spin-SU(2) symmetry. One can also define local pseudo-spins $\vec{\eta}_i=P\vec{S}_iP$, where $P$ is a Nambu transformation such that $Pc_{i\uparrow}P=c_{i\uparrow}$, $Pc_{i\downarrow}P=(-1)^i c_{i\downarrow}^\dagger$. Explicitly,
\begin{align} \eta_j^x &= \frac{1}{2}(-1)^j (c_{j\uparrow}^\dagger c_{j\downarrow}^\dagger +c_{j\downarrow}c_{j\uparrow}),\\
\eta_j^y&=\frac{1}{2}(-1)^j i(c_{j\uparrow}^\dagger c_{j\downarrow}^\dagger-c_{j\downarrow}c_{j\uparrow}),\\
\eta_j^z&= \frac{1}{2}(n_j-1).\end{align}
The three components express the local charge, or the deviation from half filling, in different manners.
On the other hand, the Hubbard term in $H$ can be rewritten as
\begin{align} H_U=\frac{U}{3}\sum_i (\vec{\eta}_i^2-\vec{S}_i^2),\label{eq:HU} \end{align} which changes sign upon the Nambu transformation.
At half filling ($\mu=0$), it can be shown that the total pseudo-spin $\sum_i\vec{\eta}_i$ commutes with the Hamiltonian $H$.
In this case, the Hamiltonian also has charge-SU(2) symmetry.
So the model is $\text{SU(2)}\times \text{SU(2)} /\text{Z}_2= \text{SO(4)}$ symmetric (with Z$_2$ the staggered particle-hole transformation) for finite $U$ \cite{yang1990so},
and $\text{SO(4)}\rtimes \text{Z}_2 =\text{O(4)}$ (with Z$_2$ the Nambu transformation defined above) for $U=0$ \cite{cai2021antiferromagnetism}. A repulsive $U$ favors the formation of local spins, or equivalently suppresses local charge deviation from half filling.

After integrating out the phonons, we obtain a retarded electron-electron interaction described by the action $\Delta S$
\begin{align*}
\Delta S=\frac{1}{2}\sum_{\av{ij}} \int\md\tau\int\md\tau' B_{ij}(\tau)\Pi(\tau-\tau')B_{ij}(\tau'),
\end{align*}
where $B_{ij}(\tau)=\sum_\sigma \left[\bar{c}_{i\sigma}(\tau) c_{j\sigma}(\tau)+{\rm c.c.}\right]$, and $\bar{c}$ and $c$ are Grassman fields.
The retarded interaction $\Pi(\tau-\tau')$ written in the frequency space is
\begin{align}
\Pi_{\nu}=-\frac{g^2}{M\omega^2}\frac{\w^2}{\nu^2+\w^2}.
\end{align}
Here $\nu$ is the bosonic Matsubara frequency, and the minus sign indicates attraction.
Note that $M\omega^2=K$ is just the spring constant for the optical phonon mode.
As usual, we write $\Pi_0=-\lambda W$, with $W=8t$ the bare bandwidth, to define the coupling strength, $\lambda=g^2/(K W)$.

Much of the effect of the SSH phonons can be anticipated and understood as follows.

\begin{figure}
	\includegraphics[width=0.45\textwidth]{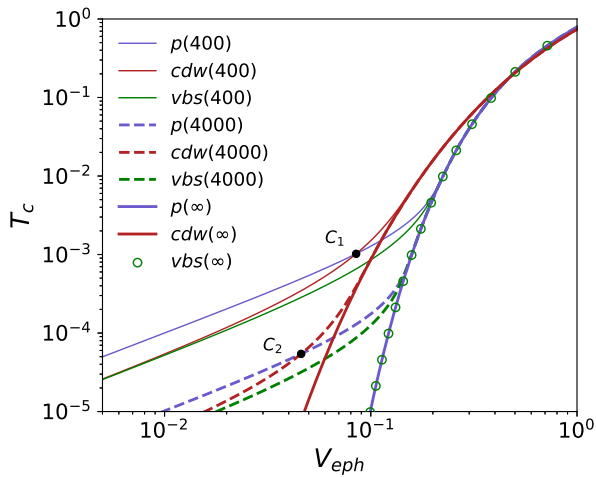}
	\caption{Mean field transition temperatures $T_c$ for Pomeranchuk, VBS and CDW  versus $V_{\rm eph}$ in $L\times L$ lattices, for $L=400$ (thin lines), $4000$ (dashed lines) and $\infty$ (thick lines). In finite-sized systems, there is a level crossing (denoted as $C_1$ and $C_2$) between the Pomeranchuk and CDW states, in favor of the Pomeranchuk state in the weak coupling limit $V_{eph}\to 0$.}
	\label{fig:mft}
\end{figure}

(i) In the adiabatic limit $\omega\rightarrow 0$, the SSH phonon acts as classical bond-wise bias fields to the electrons,
so it can obviously induce ordering of bond-wise operators. There are two types of favorable bond operators.
The first is the d-wave Pomeranchuk operator, $(\cos k_x -\cos k_y)c_{k\sigma}^\dagger c_{k\sigma}$ in the momentum space, which leads to a nematic state breaking rotation symmetry. The susceptibility for this order diverges logarithmically because of the van Hove singularity in the density of states.
The second is the doubly degenerate VBS operators, $i\sin k_{x,y}c_{k+\0Q\sigma}^\dagger c_{k\sigma}$, with an ordering momentum $\0Q=(\pi,\pi)$.
The susceptibility for this order diverges logarithmically because of the perfect nesting (while the van Hove points are suppressed by the form factor $\sin k_{x,y}$).
To gain further insight, we consider weak coupling, at which one could ignore channel overlapping and resort to mean field analysis for each ordered state separately. The transition point satisfies a generalized Stoner criterion $1=V_{\rm eph}\chi$, where $V_{\rm eph}=|\Pi_0|$ is the strength of the effective interaction, and $\chi$ is the bare susceptibility at the transition temperature $T=T_c$ for a respective order. We find that for the Pomeranchuk order,
\begin{align}
	\chi_{p}=\frac{1}{N}\sum_{k}\frac{4(\cos{k_x}-\cos{k_y})^2}{T}\cosh^{-2}\left(\frac{\varepsilon_k}{2T}\right),
\end{align}
and for the VBS state,
\begin{align}
  \chi_{vbs} =\frac{1}{N}\sum_{k}\frac{4\sin^2{k_x}}{\varepsilon_k}\tanh\left(\frac{\varepsilon_k}{2T}\right),
\end{align}
where $N$ is the number of sites, and $\varepsilon_k=2t(\cos k_x+\cos k_y)$. (We neglected the small renormalization of the uniform hopping by the coupling to the phonons).
In Fig.~\ref{fig:mft}, we plot $T_c$ versus $V_{\rm eph}$ determined by the Stoner criterion stated above. For finite-sized $400\times 400$ (thin lines) and $4000\times 4000$ (dashed lines) systems, we find the Pomeranchuk order always wins over the VBS, particularly so in the weak coupling limit. However, the two transition temperatures (or susceptibilities) become degenerate in infinite-sized systems (thick lines). This degeneracy is accidental (instead of being symmetry protected), as it is violated in finite systems.
Note the finite-size effect is weak or absent if $T_c$ is relatively higher (and hence larger than the single-particle level spacing).

(ii) In the anti-adiabatic limit $\w\rightarrow \infty$, the phonon-induced action $\Delta S$ amounts to an instantaneous interaction
\begin{align}
-\frac{V_{\rm eph}}{2}\sum_{\langle ij\rangle}B_{ij}^2
= 2V_{\rm eph}\sum_{\langle ij\rangle}(\vec{S}_i\cdot \vec{S}_j+\vec{\eta}_i\cdot\vec{\eta}_j-\frac{1}{4}),
\end{align}
where
$B_{ij}$ is now understood as a bond operator.  Clearly, this interaction favors the bond order, as well as the antiferromagnetic ordering of local spins or local pseudo-spins.
At half filling, the latter orders are related by the Nambu transformation, so that the antiferromagnetic spin order is degenerate
with the uniform s-wave pairing and CDW at momentum $\0Q$. The mean field transition point can also be written as a Stoner-like criterion, $V_{\rm eph}\chi_{cdw}=1$, with the
susceptibility (degenerate for CDW/SDW/sSC by symmetry) given by
\begin{align} \chi_{cdw}=\frac{1}{N}\sum_{k}\frac{2}{\varepsilon_k}\tanh\left(\frac{\varepsilon_k}{2T}\right),
\end{align}
which diverges double-logarithmically, because of the perfect nesting as well as the van Hove singularity. Therefore, the local spin ordering, up to the O(4) symmetry, may be more favorable in the anti-adiabatic and weak coupling limits.
One then anticipates a transition from bond-wise order at low phonon frequency to site-local order at high frequency.
The transition temperature for CDW/SDW/sSC, determined by the Stoner-like criterion, is also shown
in Fig.~\ref{fig:mft}. Note that in the same anti-adiabatic limit, the mean field transition points for the Pomeranchuk and CDW states are exactly the same as in the adiabatic limit. The comparison shows that
the CDW/SDW/sSC is favored in the thermodynamic limit, as anticipated by the double-logarithmic divergence in the susceptibility. Interestingly, in the finite-sized $400\times400$ and $4000\times4000$ systems, we always find the Pomeranchuk order wins in the weak coupling limit $V_{\rm eph}\to 0$. The crossing points $C_1$ and $C_2$ mark the level crossing between Pomeranchuk and CDW states.

(iii) Finally, from the expression of $H_U$ in Eq.\ref{eq:HU}, it is clear that a repulsive $U$ favors (suppresses) the ordering in the
site-local spin (charge) channel.
Away from half filling, the charge-SU(2) symmetry is broken, and both the van Hove singularity and perfect nesting are absent. We then expect weakening of orderings in the particle-hole channels, while the pairing susceptibility in the Cooper channel is always logarithmically divergent. We then expect pairing triggered by fluctuations in the CDW and/or SDW channels at weak coupling, and CDW/SDW/VBS at incommensurate momentum at strong coupling.

\begin{figure*}
	\includegraphics[width=0.7\textwidth]{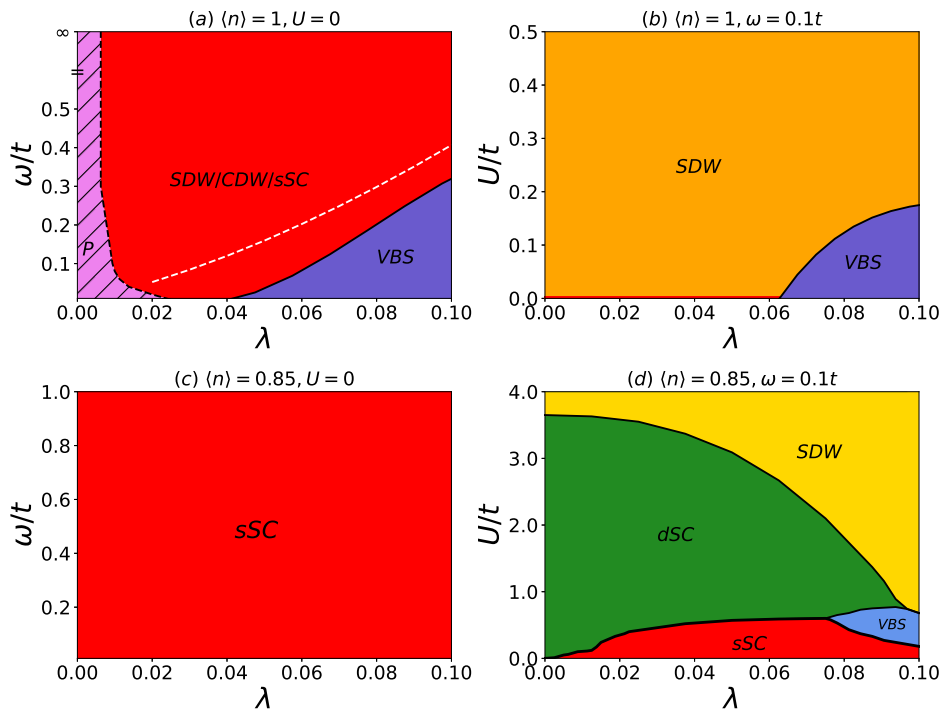}
	\caption{Phase diagram in the respective parameter space for (a) $\av{n}=1$ and $U=0$, (b)  $\av{n}=1$ and $\omega=0.1t$, (c) $\av{n}= 0.85$ and $U=0$, and (d)  $\av{n} =0.85$ and $\omega =0.1t$. In panel (a), the hatched regime indicates the Pomeranchuk instability in finite-sized systems, which shrinks in the thermodynamic limit. The white dashed line denotes the phase boundary extrapolated from the QMC result in Ref.~\onlinecite{cai2021robustness}. The other states are defined in the text. The momentum of the SDW/CDW/VBS states is $\0Q$ at half filling, and is close to $\0Q$ (slightly incommensurate) in the doped case. }
	\label{fig:phase}
\end{figure*}

The plethora of ordering tendencies in the SSH-Hubbard model discussed above must be treated on the same footing.
For this purpose, we resort to
SMFRG \cite{wang2012functional, xiang2012high, wang2013competing}. The idea of FRG is to inspect the flow of one-particle irreducible (1PI) vertices, namely, the effective interactions between quasiparticles, with respect to a decreasing infrared cutoff energy scale $\Lambda$ (the smallest fermion Matsubara frequency in our case). As in usual practice, we limit ourselves to the four-point vertices. It is understood that, by power counting, this truncation is valid, controlled and self-consistent if the instability occurs at low energy scales. We therefore limit ourselves to weak and moderate coupling regimes. The effect of EPI is taken into account through $\Pi_\nu$, which enters the flow equations as described in Ref.~\onlinecite{wang2015phonon} and the Appendix. The four-point vertex functions can be rewritten (or decomposed) as scattering matrices for all types of fermion bilinears in all collective channels, namely, the pairing channel for SC, the SDW, and CDW channels. Note that these channels are able to capture both onsite and on-bond density-density interactions. (Following the convention but at the risk of possible slight confusion, when an order parameter is concerned, we use SDW/CDW to indicate site-local orders, while the orders on bond, such as the Pomeranchuk and VBS orders, are specifically indicated, although they all belong to the general CDW channel.)
Since various decompositions are performed with respect to the same interaction vertices, the ordering tendencies in the three channels are treated on equal footing. In SMFRG, the fermion
bilinears are truncated in the internal spatial range between the two fermions within a fermion bilinear, up to a range $L_c$ that is sufficient to capture all types of potential singular scattering modes. Note that the setback distance between two bilinears is unlimited, and the dependence in this distance is resolved by the collective momentum of the bilinear. The SMFRG is strictly momentum conserving, and is asymptotically exact (up to the four-point vertices) at large $L_c$. In practice, $L_c=\sqrt{2}$ in our case is sufficient to capture all potential orders involving fermion bilinears defined on-site, on nearest-neighbor bonds and on next-nearest-neighbor bonds. This includes the local (as well as nonlocal) spins in the SDW channel; local charges, Pomeranchuk operator, and VBS operator in the CDW channel; local s-wave pairing and d-wave pairing on bonds in the SC channel, etc. The SMFRG has been applied successfully in various contexts \cite{wang2012functional, xiang2012high, wang2013competing, wang2014triplet, yang2014, wang2015phonon, lyc2017, lyc2018, wang2019theory, tang2019, Yang20201901}, and recently it is also re-interpreted as truncated-unity FRG \cite{LICHTENSTEIN2017100}. We monitor the leading (most negative) singular values $S$ of the scattering matrices in the three channels, as functions of the scale $\Lambda$ (as well as the collective momentum ${\bf q}$). The first divergence of $S$ in one out of the three channels indicates the instability of the normal state toward the formation of a long range order in that channel.
The corresponding energy scale $\Lambda_c$ gives a measure of the transition temperature $T_c$ of that order, and the eigen scattering mode (as well as the associated momentum) describes how fermion bilinears are linearly combined into the order parameter.
(Academically, if the ordering breaks a continuous symmetry in 2D and at finite temperatures, it is understood as an instability, quasi-long range order, or the onset of strong correlations at finite temperatures in 2D, in view of the Mermin-Wagner theorem.)
We refer to appendix \ref{sec:appendixA} and Refs.~\onlinecite{wang2013competing, wang2015phonon, tang2019} for more technical details.

\section{RESULTS AND DISCUSSIONS} \label{sec:results}

\begin{figure*}
	\includegraphics[width=0.7\textwidth]{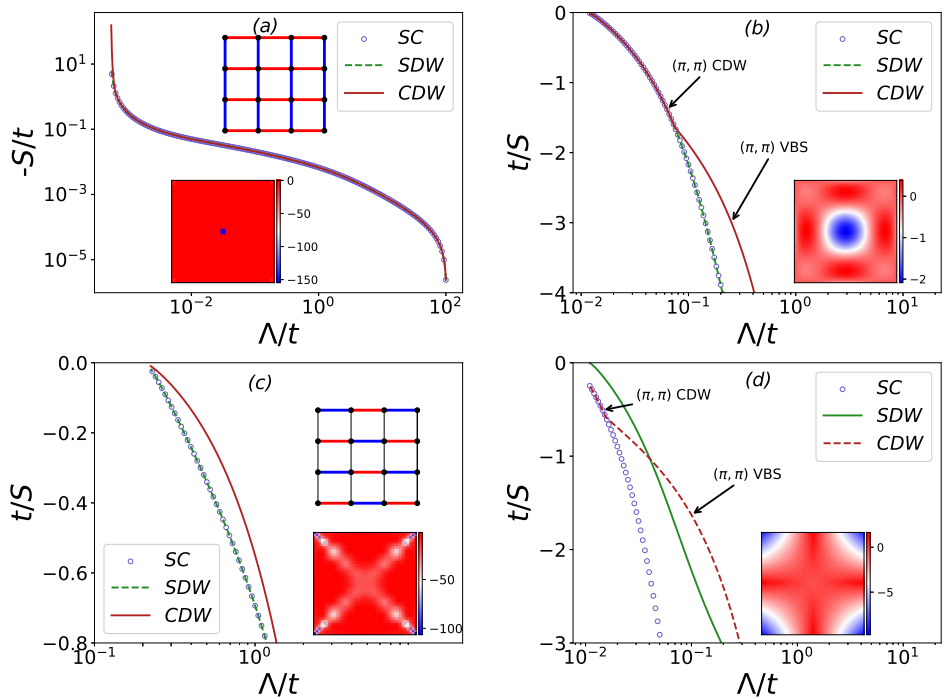}
	\caption{ FRG flows of the leading singular values $S$ in the SC, SDW and CDW channels at half-filling $\av{n}=1$, in four representative cases: (a) $U=0$, $\omega=\infty$, and $\lambda=0.005$. The lower inset shows $S(\0q)$ in the CDW channel, and the upper inset shows the leading eigen mode, the Pomeranchuk state in real space;
	(b) $U=0$, $\omega=0.3t$, and $\lambda=0.025$. The inset shows $S(\0q)$ in the pairing channel. {The arrows show there is a level crossing from bond-bilinear to site-bilinear in the CDW channel.} At the final stage, the leading eigenvalues in the SDW, CDW, and SC channels are degenerate;
	(c) $U=0$, $\omega=0.1t$, and $\lambda=0.075$. The lower inset shows $S(\0q)$ in the CDW channel, and the upper inset shows the leading eigen mode, the VBS state in real space;
	(d) $U=0.1t$, $\omega=0.1t$, and $\lambda=0.025$. The inset shows $S(\0q)$ in the SDW channel.}
	\label{fig:flow1}
\end{figure*}
We performed systematic calculations using SMFRG to obtain the phase diagrams in Fig.~\ref{fig:phase}. For $\av{n}=1$ and $U=0$ in Fig.~\ref{fig:phase}(a), we find the nematic Pomeranchuk instability in the weak coupling limit, as a result of finite size effect, consistent with the mean field analysis. This phase shrinks (in $\lambda$) with denser momentum grid used in the SMFRG calculations, and should disappear in the thermodynamic limit. On the other hand, we obtain the degenerate SDW/CDW/sSC states for larger $\lambda$ and larger phonon frequency, while the VBS state appears at smaller phonon frequency. The phase boundary between SDW/CDW/sSC and VBS agrees very well with that (the white dashed line) extrapolated from the QMC results in Ref.~\onlinecite{cai2021robustness}. (Note the slight deviation is reasonable, as the extrapolation is not necessarily rigorous to the weak and moderate coupling suitable for FRG calculations). For $\av{n}=1$ and $\omega=0.1t$ in Fig.~\ref{fig:phase}(b), we find the SDW state for larger $U$ and VBS state for larger $\lambda$. For $\av{n}=0.85$ and $U=0$ in Fig.~\ref{fig:phase}(c), the only instability is sSC for all nonzero phonon frequency and coupling $\lambda$. For $\av{n}=0.85$ and $\omega=0.1t$ in Fig.~\ref{fig:phase}(d), we obtain the sSC, dSC and incommensurate SDW as $U$ increases for a small up to moderate $\lambda$. As $\lambda$ is also large, we find a small window of incommensurate VBS.
In the following we substantiate and discuss the details of the FRG flow at typical points in each of the phase diagrams.

We note that in our FRG calculations, we only consider phonons at finite frequencies. The reason is a zero-frequency Holstein-like phonon mode is already unstable, and becomes more so by the polarization effect from the coupling to the electrons. The physics is more straightforwardly understood by the mean field theory, as we discussed, which becomes actually exact at zero temperature (when thermal fluctuations of the classical phonon fields are absent).

\subsection{$\av{n}=1$ and $U=0$} \label{subsecA}

Fig.~\ref{fig:flow1}(a) shows the FRG flow of the leading singular values $S$ in the three channels for $\omega = \infty$ and $\lambda = 0.005$.
Here, a log-log plot is adopted for clarity.
At high energy scales, the three channels are degenerate, corresponding to the degenerate SDW/CDW/sSC state.
But as $\Lambda$ decreases, the leading eigen mode in the CDW channel changes from on-site bilinear at collective momentum $\0Q=(\pi,\pi)$,
to on-bond bilinear $\sum_{\sigma\delta} \xi_\delta c_{i\sigma}^\dag c_{i+\delta,\sigma}$
at $\0q=(0,0)$, with $\xi_\delta=1/2$ for $\delta=\pm \hat{x}$ and $\xi_\delta=-1/2$ for $\delta=\pm \hat{y}$. In momentum space, this reads $\sum_{\sigma}(\cos k_x-\cos k_y)c_{k\sigma}^\dagger c_{k\sigma}$, which is just the Pomeranchuk operator.
This mode then grows up quickly and diverges first. The emerging Pomeranchuk state causes nematic modulation of the hopping integrals along $x$ and $y$ directions, as shown in the inset in Fig.~\ref{fig:flow1}(a), where the red (blue) bond indicates strong (weak) hopping.
The Fermi surface is changed such that the van Hove singularity (VHS) points at $(\pi,0)$ and $(0,\pi)$ are avoided.
However, as shown in Sec.~\ref{sec:model}, the existence of Pomeranchuk state at weak coupling is caused by the finite size effect. We checked that the Pomeranchuk regime shrinks (in $\lambda$) as we use denser momentum grid in the one-loop integrations for FRG. On the other hand, at stronger coupling and small phonon frequency, we find the VBS state wins over the Pomeranchuk state, breaking the accidental mean field degeneracy, as a result of the competition between interactions responsible for such orders that are treated equally in our FRG.

Fig.~\ref{fig:flow1}(b) shows the FRG flow for $\w=0.3t$ and $\lambda=0.025$.
At high energy scales, the SC and SDW channels are degenerate, but are split from the CDW channel.
This is because the leading scattering mode in the CDW channel is on-bond at high $\Lambda$ and thus is naturally different to the on-site modes of SDW/sSC at this stage.
As the flow continues toward lower energy scales, the three channels merge and diverge simultaneously.
We checked that here the leading eigen scattering modes are all local bilinears.
The inset shows $S(\0q)$ in the SC channel as a function of the momentum $\0q$, from which the strong negative peak at $\0q=0$ indicates the Cooper pairing at zero center-of-mass momentum.
Meanwhile, the other two channels are found (not shown) to peak at $\0q=\0Q$, corresponding to the site-local SDW and CDW orders.
The existence of such a degenerate SDW/CDW/sSC state is due to the underlying O(4) symmetry, and is consistent with recent QMC results \cite{feldbacher2003coexistence, cai2021antiferromagnetism, cai2021robustness, xing2021quantum, gotze}.

\begin{figure*}
	\includegraphics[width=0.7\textwidth]{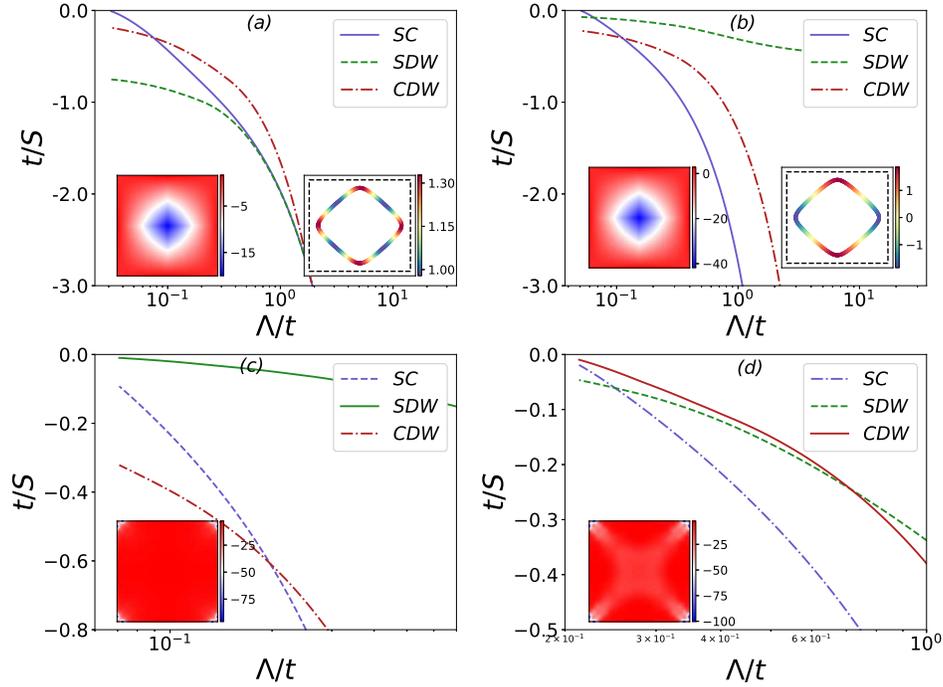}
	\caption{ FRG flows of the leading singular values $S$ in the SC, SDW and CDW channels at doping level $\av{n}=0.85$, in four representative cases:
	(a) $U=0$, $\omega=0.1t$, and $\lambda=0.05$. The left inset shows $S(\0q)$ in the diverging SC channel, and the right inset shows the gap function on the Fermi surface;
	(b) $U=2t$, $\omega=0.1t$, and $\lambda=0.05$. The insets are defined similarly to that in panel (a);
	(c) $U=3.8t$, $\omega=0.1t$, and $\lambda=0.0125$. The inset shows $S(\0q)$ in the diverging SDW channel;
	(d) $U=0.5t$, $\omega=0.1t$, and $\lambda=0.0875$. The inset shows $S(\0q)$ in the diverging CDW channel.}
	\label{fig:flow2}
\end{figure*}

In Fig.~\ref{fig:flow1}(c), we plot the FRG flow for $\w=0.1t$ and $\lambda=0.075$.
During the entire flow, the leading channel is CDW, although the other two channels also catch up quickly at low energy scales, indicating the SSH EPI favors both on-bond VBS and on-site SDW/CDW/sSC orders, and the VBS only wins narrowly.
Here, for the VBS, the ordering momentum $\0q$ is found to be $\0Q$ from the negative peak in $S(\0q)$ shown as the lower inset of Fig.~\ref{fig:flow1}(c).
We find two degenerate scattering modes at this momentum in the CDW channel, $i \sin k_{x,y} c_{k+Q}^\dagger c_k$ in momentum space. The form factor has the $p_x$ or $p_y$ symmetry, and the degeneracy is protected by the little group at $\0Q$ (which is the same as the point group).
The real-space pattern of this state, up to the two-fold symmetry, is schematically shown in the upper inset of Fig.~\ref{fig:flow1}(c), where red/blue color stands for strong/weak bond.
In the ordered state, the two degenerate modes may be linearly recombined to minimize the energy.

Our systematic results for this subsection are summarized in Fig.~\ref{fig:phase}(a).
We find three competing phases, namely, the nematic phase from Pomeranchuk instability, the degenerate SDW/CDW/sSC phase, and the VBS phase.
The Pomeranchuk phase is caused by the finite size effect and should shrink toward $\lambda=0$. The phase boundary between SDW/CDW/sSC and VBS is qualitatively
in agreement with the QMC results \cite{cai2021robustness} extrapolated to the weak coupling regime (white dashed line).
We should point out that the SDW/CDW/sSC state in our case is degenerate exactly, as our FRG respects all relevant symmetries exactly. However, a spontaneous symmetry breaking may happen in the ordered state, as claimed in QMC \cite{cai2021antiferromagnetism}, and may be sensitive to the implementation details of QMC.

\subsection{$\av{n}=1$ and $U>0$} \label{subsecB}

The FRG flow for $\w=0.1t$, $\lambda=0.025$ and $U=0.1t$ is presented in Fig.~\ref{fig:flow1}(d).
The leading channel at high energy scales is CDW and its scattering mode corresponds to the VBS order, caused by the strong EPI relative to $U$.
As the energy scale is lowered, the SDW channel grows up quickly and diverges first.
Its eigen scattering mode is site-local and the collective momentum is $\0Q$ from the plot of $S(\0q)$ (inset) in Fig.~\ref{fig:flow1}(d). This is just the antiferromagnetic SDW state.
In comparison to the case of $U=0$ with degenerate SDW/CDW/sSC state, here, a repulsive $U$ breaks the Z$_2$ symmetry and selects the SDW as the dominant instability.
In addition, due to the remaining SO(4) symmetry, the CDW and sSC channels are still degenerate at low energy scales.

For the phonon frequency $\w=0.1t$, we obtain the phase diagram in the $(\lambda,U)$ space in Fig.~\ref{fig:phase}(b).
A large positive $U$ favors SDW, while
a large enough $\lambda$ drives VBS. Since the O(4) symmetry reduces to SO(4) by a finite $U$, the SDW state has no longer degenerate counter part in the charge channel.

\subsection{$\av{n}=0.85$ and $U=0$} \label{subsecC}

Next we examine the effect of finite doping.
We consider the filling level $\av{n}=0.85$ as a typical example.
The FRG flow for $\omega=0.1t$ and $\lambda=0.05$ is shown in Fig.~\ref{fig:flow2}(a).
At high energy scales, the CDW channel dominates.
With decreasing $\Lambda$, the SC channel is triggered and eventually diverges  first.
This behavior indicates the superconductivity is related to charge fluctuations.
The leading eigen mode of the SC channel is dominated by onsite electron pair, with center-of-mass momentum $\0q=0$, as seen from the left inset. The right inset shows the pairing function on the Fermi surface. The slight modulation along the Fermi surface comes from the subdominant off-site pairs in the eigen mode.

The phase diagram (for this subsection) in the $(\lambda,\w)$ space is shown in Fig.~\ref{fig:phase}(c), with the sSC phase alone up to $\lambda=0.1$.
This is reasonable, since the bare Cooper pairing susceptibility (at $\0q=0$) always enjoys logarithmic divergence as long as the normal state is time-reversal invariant and/or inversion symmetric. Instead, in the particle-hole channels, the divergence is lost once the Fermi surface nesting and van Hove singularity at zero energy are absent upon finite doping.

\subsection{$\av{n}=0.85$ and $U>0$} \label{subsecD}

The FRG flow for $\w=0.1t$, $\lambda=0.05$ and $U=2t$ is shown in Fig.~\ref{fig:flow2}(b).
At high energy scales, the leading channel is SDW, suggesting strong spin fluctuations caused by Hubbard $U$.
With decreasing $\Lambda$, the SC channel is triggered and diverges.
The pairing eigen mode is at momentum $\0q=0$ as seen from $S(\0q)$ in the left inset of Fig.~\ref{fig:flow2}(b).
The right inset plots the pairing function evaluated on the Fermi surface, which clearly shows the d-wave symmetry. Therefore the system develops dSC, which can be related to the  spin fluctuations.

The FRG flow for $\w=0.1t$, $\lambda=0.0125$ and $U=3.8t$ is shown in Fig.~\ref{fig:flow2}(c).
During the entire flow, the SDW channel is the leading one.
The scattering mode is found to be local, and the collective momentum $\0Q_1$ is incommensurate and near $\0Q$, as can be seen from $S(\0q)$ in the inset of Fig.~\ref{fig:flow2}(c).
The incommensurate ordering momentum is caused by the lacking of perfect nesting at finite doping.

The FRG flow for $\w=0.1t$, $\lambda=0.0875$ and $U=0.5t$ is presented in Fig.~\ref{fig:flow2}(d).
Starting from higher energy scale, the SDW and CDW channels are stronger than the SC channel, and the CDW channel grows faster at low energy scales and diverges first.
The eigen mode is found to be a bond-wise order similar to the VBS at half filling, except that the ordering momentum $\0Q_1$ is incommensurate and near $\0Q$, see the inset for $S(\0q)$ in Fig.~\ref{fig:flow2}(d).
Note that at low energy scales, the SC channel also becomes strong, as a result of the channel overlaps in the various channels.

By systematic calculations, we obtain the phase diagram, Fig.~\ref{fig:phase}(d), for $\av{n}=0.85$ and $\w=0.1t$.
For very small $U\ll \lambda W$, the ground state is always sSC. However, this sSC state is suppressed quickly as $U$ increases, and yields to the dSC state and subsequently the incommensurate SDW state for smaller $\lambda$. On the other hand, we find the incommensurate VBS state in a narrow window centered at $\lambda\sim 0.09$ and $U\sim 0.5$.
The reason that $U$ can tune the balance between the sSC and dSC is because the SSH phonons  can trigger both CDW and SDW fluctuations, while a repulsive $U$ enhances (suppresses) the SDW (CDW) fluctuations. The CDW fluctuations are related to sSC, while the SDW fluctuations are related to dSC (in the present model).

\section{SUMMARY} \label{sec:summary}
In this work, we have studied the 2D SSH-Hubbard model on the square lattice within SMFRG.
The interplay of Hubbard $U$, EPI strength $\lambda$ and phonon frequency $\w$ leads to rich phase diagrams.
(i) At half-filling and $U=0$, we obtain degenerate SDW/CDW/sSC state at higher phonon frequency and VBS state at lower frequency and larger $\lambda$,
in agreement with QMC.
At weak coupling, we obtained the nematic Pomeranchuk instability in finite-size system, which should disappear in the thermodynamic limit. This finite-size effect remains even for a very large lattice (as the mean field analysis shows), and may pose a challenge for QMC in this regime. (ii) Upon finite doping, we find the sSC alone for $U=0$,
while an increasing $U$ drives the system into the dSC and incommensurate SDW successively. Besides, the incommensurate VBS state is realized in a narrow window of moderate $\lambda$ and $U$. These results at finite doping are beyond the present reach of QMC because of the minus sign problem.

\begin{acknowledgements}
D.W. thanks Z.-X. Liu for helpful discussions on the symmetries.
This work is supported by the National Natural Science Foundation of China (under Grant No. 11874205, No. 12274205 and No. 11574134).
\end{acknowledgements}


\appendix
\section{SMFRG with retarded interactions} \label{sec:appendixA}
The method of functional renormalization group (FRG) has been well explained in many reviews \cite{Berges_PR_2002,Metzner_RMP_2012,Dupuis_PR_2021} and textbooks \cite{Kopietz__2010}. In this appendix, we mainly focus on one of its realizations, called singular mode FRG (SMFRG), in particular with phonon induced retarded interactions as employed in this work.

\begin{figure}
	\includegraphics[width=0.45\textwidth]{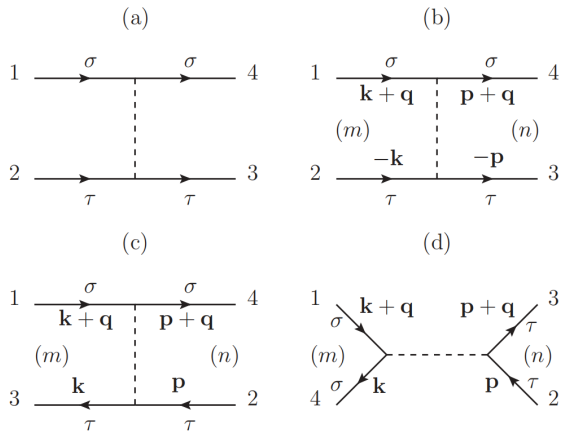}
	\caption{ A generic 4-point 1PI vertex (a) can be rearranged into the pairing (P), crossing (C) and direct (D) channels as shown in (b)-(d), respectively.
	The momentum $\mathbf{k,q,p}$ are explicitly shown for clarity. The spins ($\sigma$ and $\tau$) are conserved during fermion propagation for the spin SU(2) symmetric case. The labels $m$ and $n$	denote fermion bilinears. }
	\label{fig:vertex}
\end{figure}

\subsection{FRG flows}
\begin{figure}
	\includegraphics[width=0.45\textwidth]{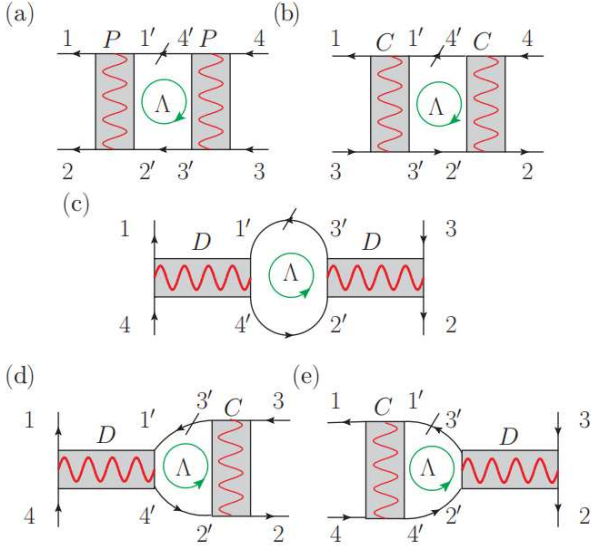}
	\caption{ One-loop contributions to $\partial\Gamma_{1234}/\partial\Lambda$. The gray bar and wavy line denote the contribution from $\Gamma$ and $\Pi_{\nu}$ respectively. They are added up in the calculation. The slash denotes the single-scale propagator and can be put on either
    one of the fermion lines within the loop. The directed-circle indicates circulation of frequency within the loop, and
    $\Lambda$ is the runing energy scale. Note that $\Pi$ enters at Matsubara frequency $\nu=\Lambda$ (thin wavy lines) in P and C channels, while it does at $\nu=0$ (thick wavy lines) in the D channel.}
	\label{fig:1loop}
\end{figure}

In our SMFRG, we study the RG flow of the 4-point 1PI vertices $\Gamma_{1234}$ appearing in the effective interaction  $\frac{1}{2}\sum_{1,2,3,4}\psi^\dagger_1\psi^\dagger_2\Gamma_{1234}\psi_3\psi_4$. Here $\psi$ is the fermion field, its
subscript labels the one-particle state, and spin conservation is assumed implicitly for spin-SU(2) symmetric systems.
The vertex can be rewritten as scattering matrices in the three Mandelstam channels by choosing different fermion bilinears. This includes the pairing ($P$), crossing ($C$), and direct ($D$) channels, and they are related as
\begin{align}
\Gamma_{1234}=P_{12;43}=C_{13;42}=D_{14;32}.
\end{align}
Correspondingly, the collective momentum $\0q$ are identified differently as shown in Fig.~\ref{fig:vertex} explicitly.
As the energy scale $\Lambda$ runs, $\Gamma_{1234}$ flows as
\begin{align} \label{eq:flow}
\frac{\partial\Gamma_{1234}}{\partial\Lambda} &= [P\chi_{pp}P]_{12;43}+[C\chi_{ph}C]_{13;42} \nn\\
&+[-2D\chi_{ph}D+D\chi_{ph}C+C\chi_{ph}D]_{14;32} ,
\end{align}
where products imply convolutions, and $\chi_{pp}$ and $\chi_{ph}$ are differential susceptibilities given by
\begin{align}
	[\chi_{pp}]_{12;34} &= \frac{1}{2\pi}\left[ G_{14}(i\Lambda)G_{23}(-i\Lambda)+
	(\Lambda \to -\Lambda) \right] \\
	[\chi_{ph}]_{12;34} &= \frac{1}{2\pi}\left[ G_{14}(i\Lambda)G_{32}(i\Lambda)+
	(\Lambda \to -\Lambda) \right],
\end{align}
where $G_{12}(i\Lambda)$ is the normal state Green's function with imaginary frequency $i\Lambda$.
As usual, we have neglected the six-th and higher order vertices, which are RG irrelevant \cite{Metzner_RMP_2012}. The Fenyman diagrams contributing to the flow equation are illustrated in Fig.\ref{fig:1loop}. Note that after $\Gamma$ is updated during the FRG flow, it is redistributed into the three Mandelstam channels as in Fig.\ref{fig:vertex}. This is why FRG can treat interactions in all channels on equal footing.

\subsection{Including phonon-meadiated interactions}
The phonon-mediated interaction can be included as a part of $\Gamma_{1234}$, with $\psi_1$ and $\psi_4$ on a bond, and $\psi_2$ and $\psi_3$ on the same bond, together with the frequency dependence in $\Pi_\nu$.
In principle, Eq.~\ref{eq:flow} can be directly applied to the total interaction including the retarded one, by keeping the full frequency dependence. However, the frequency-dependence in the FRG-generated correction to the four-point vertices can be argued to be RG irrelevant\cite{Kopietz__2010, Metzner_RMP_2012}. In this spirit, we separate the total vertex $\Gamma$ into an instantaneous part $\Gamma^I$ and a retarded one $\Gamma^R$,
\begin{align}
\Gamma=\Gamma^I+\Gamma^R.
\end{align}
Since we take the FRG-corrected part as instantaneous, the retarded part is always given by $\Gamma^R=\Pi_\nu$ for the associated fermions. The initial value of $\Gamma^I$ at $\Lambda=\infty$ is given by the Hubbard $U$. For brevity, we will keep using the notations $\Gamma$, $P$, $C$, and $D$ (without superscript ``$I$'') for the instantaneous part.
The Feynman diagrams contributing to the flow of $\Gamma$ is illustrated in Fig.\ref{fig:1loop}, where the gray bars are the instantaneous vertices, and the wavy lines are from the retarded kernel suitably added to the instantaneous part. Explicitly, the flow equation can be written as
\begin{align} \label{eq:ph-flow}
\frac{\partial\Gamma_{1234}^I}{\partial\Lambda} &= [\+P\chi_{pp}\+P]_{12;43}+[\+C\chi_{ph}\+C]_{13;42} \nn\\
&+[-2\+D\chi_{ph}\+D + \+D\chi_{ph}\+C + \+C\chi_{ph}\+D]_{14;32} ,
\end{align}
where $\+P=P+[\Pi_\Lambda]_P$, $\+C=C+[\Pi_\Lambda]_C$, $\+D=D+[\Pi_0]_D$, with $[\Pi_\nu]_{P,C,D}$ the phonon-induced interaction projected in the respective channels (or associated to the desired fermion bilinears).
Note that the Matsubara frequency of $\Pi_\nu$ is $\Lambda$ in $\+P$, $\+C$, and $0$ in $\+D$, as a result of frequency conservation when the external fermions are all set at zero frequency in Fig.\ref{fig:1loop}.

\begin{figure}[h!]
	\includegraphics[width=0.4\textwidth]{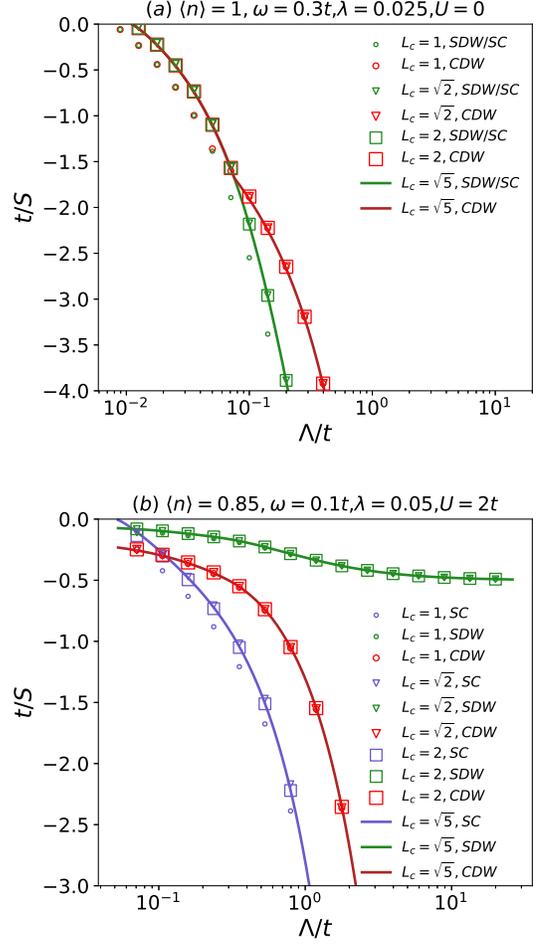}
	\caption{SMFRG results for various choices of the truncation length $L_c$ in both undoped (a) and doped (b) cases, with the same parameters as in Fig.~\ref{fig:flow1}(b) and Fig.~\ref{fig:flow2}(b), respectively.}
	\label{fig:check}
\end{figure}

\subsection{Truncated fermion bilinears and singular mode analysis}

Without truncation in the fermion bilinears, $P$, $C$ and $D$ are equivalent to, or different nick names of, $\Gamma$. Since the number of fermion bilinears are infinite, in practice, one has to truncate the bilinears (modulo the center of mass position) into a finite set. Mathematically, this can also be understood as a truncation of the completeness condition, or truncated unity \cite{LICHTENSTEIN2017100}. To capture ordering tendencies for short-range operators, which are arguably the most usual and important, we limit the internal range between the two fermions within a bilinear up to a truncation length $L_c$. This means that in Fig.\ref{fig:vertex}, we require $|\0r_1-\0r_2|\le L_c$ and $|\0r_3-\0r_4|\le L_c$ for $P$,  $|\0r_1-\0r_3|\le L_c$ and $|\0r_2-\0r_4|\le L_c$ for $C$, and $|\0r_1-\0r_4|\le L_c$ and $|\0r_2-\0r_3|\le L_c$ for $D$.
These vertices, working as scattering amplitude between important fermion bilinears, are most likely driven by FRG into singular scattering modes, as the naming of SMFRG implies. In momentum space, these vertices become matrices $P(\0q)$, $C(\0q)$ and $D(\0q)$ in the bilinear basis, at momentum $\0q$ in the respective channel.

In the spin-SU(2) symmetric case, the effective interactions in the SC, SDW and CDW channels are given by
\begin{align}
V^{\rm SC}=P,\ \ V^{\rm SDW}=-C,\ \ V^{\rm CDW}=2D-C.
\end{align}
In a specific channel, the interaction can be decomposed as
\begin{align}
V_{mn}(\0q)=\sum_{\alpha} \phi_{\alpha}(m) S_{\alpha}(\0q) \phi_{\alpha}^*(n),
\end{align}
where $m$ and $n$ label the fermion bilinear, $S_{\alpha}(\0q)$ is the singular value for the eigen scattering mode $\phi_{\alpha}$ in the bilinear basis.
The leading (most negative) singular value $S$ (out of all $\0q$'s) in each channel is monitored with decreasing $\Lambda$. The first divergence indicates the instability to an ordered state described by the corresponding eigen scattering mode momentum $\0q$. In practice, we choose $S=100t$ as the criterion of divergence.

In the main text, we choose the truncation length $L_c=\sqrt{2}$ for fermion bilinears, which is sufficient to capture all the potential orders under concern. Here, we check the convergence of our results with respect to $L_c$, as shown in Fig.~\ref{fig:check}.
With the same parameters as in Fig.~\ref{fig:flow1}(b) and Fig.~\ref{fig:flow2}(b), the leading singular values in all three channels saturate already as $L_c\ge\sqrt{2}$, justifying the truncation we applied in the main text.

\bibliography{reference-ssh-note}

\end{document}